\documentclass[
aps,%
12pt,%
final,%
notitlepage,%
oneside,%
onecolumn,%
nobibnotes,%
nofootinbib,
superscriptaddress,%
noshowpacs,%
centertags]%
{revtex4}
\RequirePackage{graphicx}

\def\refitem#1{\relax}

\begin{document}
\title{Methods to study event-by-event fluctuations in the NA61/SHINE experiment at the CERN SPS}

\author{\firstname{T.} \surname{Cetner}}
\email{Tomasz.Cetner@cern.ch}
\affiliation{Faculty of Physics, Warsaw University of Technology, Poland}

\author{\firstname{K.} \surname{Grebieszkow} (for the NA61 Collaboration)}
\email{kperl@if.pw.edu.pl}
\affiliation{Faculty of Physics, Warsaw University of Technology, Poland}


\begin{abstract}

Theoretical calculations locate the critical point of strongly interacting matter (CP) at energies
accessible at the CERN SPS. Event-by-event transverse momentum and multiplicity fluctuations are considered as one of the most important tools to search for the CP. Pilot studies of the energy dependence and the system size dependence of both $p_T$ and multiplicity fluctuations
were performed by the NA49 experiment.
The NA61/SHINE ion program is a continuation of these efforts. 
After briefly recalling the essential NA49 results on
fluctuations we will discuss the technical methods (removing Non-Target interactions) which we plan to
apply for future transverse momentum and multiplicity fluctuation analyses.

\end{abstract}

\maketitle

\section{Introduction}

The Super Proton Synchrotron (SPS) at CERN covers one of the most 
interesting regions of the QCD phase diagram $(T - \mu_B)$. On the one 
hand there are indications (pion production, the kaon to pion ratio, slopes of transverse mass spectra) that the energy threshold for deconfinement 
is reached already at low SPS energies \cite{kpi_paper}. On the other hand
lattice QCD calculations locate the critical point of strongly
interacting matter in the SPS energy range ($T^{CP}=162 \pm 2 MeV$, $\mu_B^{CP}=360 \pm 40 MeV$) \cite{fodor_latt_2004}.

Fluctuations and correlations may serve as a signature of the onset of deconfinement (close to the phase transition the Equation of State changes rapidly which can impact energy dependence of fluctuations), and can  
can help to locate the critical point of strongly interacting matter
(in analogy to critical opalescence we expect enlarged fluctuations close to the CP). For strongly interacting matter the maximum of CP signal is expected when freeze-out happens close to the critical point. The position of the chemical freeze-out point in the $(T - \mu_B)$ diagram can be varied by changing the energy and the size of the colliding system.

The NA49 experiment \cite{na49_nim} at the CERN SPS studied the energy dependence (beam energies 20$A$-158$A$ GeV) and the system size dependence 
($p+p$, $C+C$, $Si+Si$, and $Pb+Pb$ at the highest SPS energy)
of event-by-event transverse momentum and multiplicity fluctuations. 
There are no indications of the CP in the energy dependence of multiplicity and mean $p_T$ fluctuations in central Pb+Pb collisions. However,
an intriguing non-monotonic dependence of both $p_T$ and multiplicity fluctuations was observed for the system size dependence at the top SPS energy \cite{kg_qm09}. Those results are consistent with a critical point located at the chemical freeze-out point of $p+p$ interactions at 158$A$ GeV.
Quite recently a non-monotonic behavior of event-by-event azimuthal angle fluctuations was also observed by NA49 \cite{HQproc}.

\section{NA61/SHINE experiment}
NA61/SHINE \cite{tobczo} is a fixed target heavy-ion experiment at the CERN SPS. Its ion program is a continuation of NA49 efforts. The main components of the detector, inherited from NA49, are four large volume time projection chambers (TPC). The Vertex TPCs (VTPC-1 and VTPC-2), are located in the magnetic field of two super-conducting dipole magnets. Two other TPCs (MTPC-L and MTPC-R) are positioned downstream of the magnets symmetrically to the beam
line. Furthermore, the setup includes Time of Flight (TOF) walls and a Particle Spectator Detector (PSD). The later will be used for a very precise centrality determination (measurement of the energy of the projectile spectator nucleons). The excellent PSD resolution (about one nucleon in the studied energy range) is crucial especially for the analysis of multiplicity
fluctuations.

\begin{figure}[h]
    \centering
\vspace{-1cm}
        \includegraphics[width=0.7\textwidth]{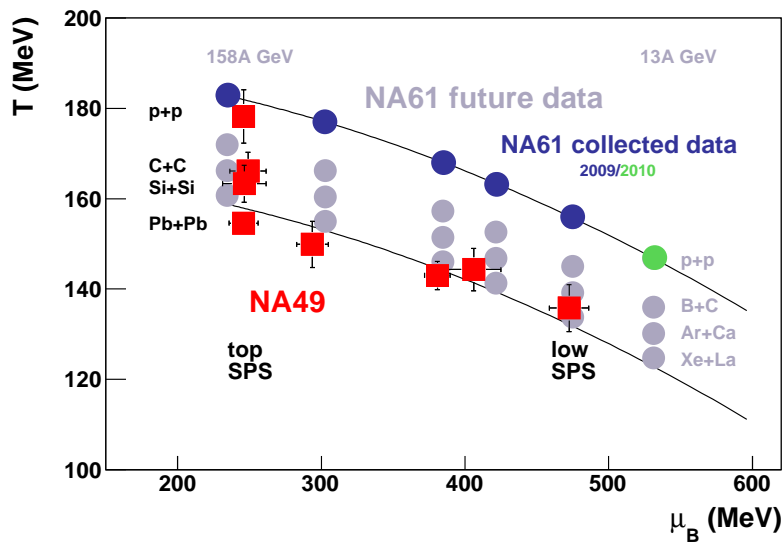}
\vspace{-0.5cm}
\caption{Two-dimensional scan of the phase diagram at SPS energies. Estimated (NA49 results Ref.~\cite{beccat}) and expected (NA61) chemical freeze-out points.}
\label{na61points}
\end{figure}

In the NA61/SHINE experiment hadron production in $p+p$, $p+A$, $h+A$, and
$A+A$ reactions at various energies will be analyzed. A broad experimental
program is planned: search for the critical point, study of the
properties of the onset of deconfinement, high $p_T$ physics, and analysis of hadron spectra for the T2K neutrino experiment and for the Pierre Auger
Observatory and KASCADE cosmic-ray experiments. Within the NA61/SHINE ion
program we plan, for the first time in history, to perform a 2D scan
with system size and energy. The data on $p+p$ (2009-2010(11)), $^{11}B+\,^{12}C$ (2011), $^{40}Ar+\,^{40}Ca$ (2013), and $^{129}Xe+\,^{139}La$ (2014)
will allow to cover a broad range of the phase diagram (see
Fig.~\ref{na61points}).

The 2009 data taking period can be considered the beginning of the ion program in NA61. For each beam momentum (20, 31, 40, 80, and 158 GeV/c) we registered several million $p+p$ events. Additionally, a test sample of $p+p$ interactions at 13 GeV/c was recorded in 2010.

\section{Extraction of Non-Target interactions}

For fixed target experiments such as NA61/SHINE, there is always an uncertainty whether the interaction occurs in the target or in some material that surrounds it. In NA61 the problem mostly concerns $p+p$ interactions where the target is 20~$cm$ long liquid hydrogen (LH) cylinder, and contamination may originate for example from collisions with mylar windows of the LH cylinder. 
There are numerous procedures designed to minimize the number of unwanted, Non-Target, interactions within data used for physics analysis. 
This includes a fine-tuned interaction trigger system as well as various quality criteria for event selection.
With this approach one can select a group of events with the highest probability of coming from interactions on the target.
However, for some analyses this may not be the most effective procedure. 
Event-by-event fluctuation measures can be sensitive to inclusion of interactions on different nuclei, in particular if this means additional $p+A$ interactions within the $p+p$ sample. Moreover, high statistics needed for those analyses can be drastically lowered by strict event selection criteria.

An opposite approach is to use a more contaminated sample for analysis, but correct the results using data acquired from interactions on Non-Target material. Such a procedure is followed in NA61, where repeatedly during the data taking period the target material is removed. For interactions on protons the liquid hydrogen container is emptied (Target Empty), for interactions on solid targets, such as carbon, the target is taken out of the beam (Target Out). The normal experiment setup is called Target Full or Target In. 

The Target Full data sample is contaminated with Non-Target interactions. 
We define $\alpha$ as the fraction of Target interactions within the Full data sample:
\begin{eqnarray}
\alpha = \frac{n^F_T}{n^F}
\end{eqnarray}
where $n^F_T$ is the number of Target interactions within Full data sample and $n^F$ is the total number of interactions within Full data sample.
For an event variable $W$ we can write
\begin{eqnarray}
\langle W \rangle_T^F =  \frac{1}{\alpha} \; [\langle W \rangle^F -  (1-\alpha) \langle W \rangle_{NT}^F]
\label{emTF}
\end{eqnarray}
where $\langle W \rangle_T^F$ is what we would like to measure in the end: event mean value for Target interactions within the Full data sample.
$\langle W \rangle^F$ is what we directly obtain from the Full data sample: the event mean value for the whole Full data sample. 
$\langle W \rangle_{NT}^F$ is what we want to correct for: event mean value for Non-Target interactions within the Full data sample.

In order to use Empty Target data, we assume that event mean values for Non-Target interactions are the same within the Full and Empty data samples: $\langle W \rangle_{NT}^F =\langle W \rangle_{NT}^E $.
Additionally, the Empty data sample consists only of Non-Target interactions $\langle W \rangle_{NT}^E = \langle W \rangle^E$,  and Target interactions are only in the Full data sample $\langle W \rangle_{T}^F = \langle W \rangle_T$.
This allows us to rewrite equation (\ref{emTF}) using quantities that are derived directly from the acquired data samples:
\begin{eqnarray}
\langle W \rangle_T =  \frac{1}{\alpha} \; [\langle W \rangle^F -  (1-\alpha) \langle W \rangle^E] 
\label{emT}
\end{eqnarray}
Equation (\ref{emT}) is the main formula allowing to calculate the event mean value for interactions on the Target.

\begin{figure}[h]
    \centering
        \includegraphics[width=0.9\textwidth]{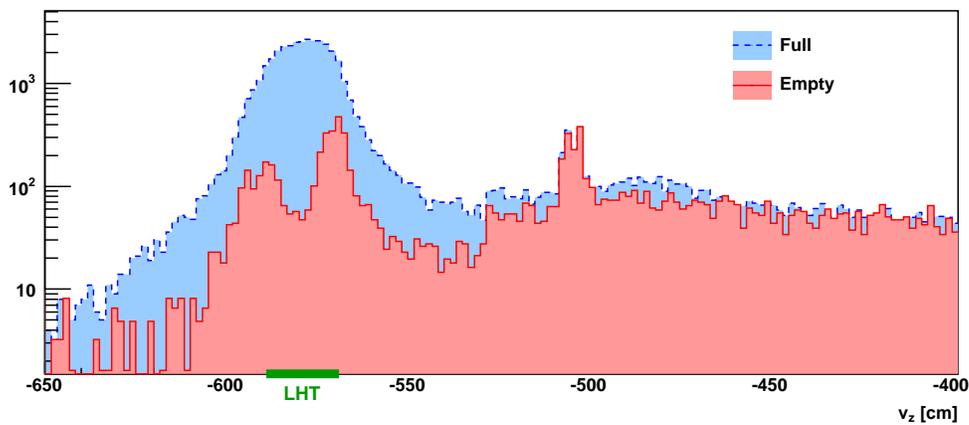}
\vspace{-0.5cm}
\caption{Distributions of vertex z position for $p+p$ collisions at beam momentum 40 GeV/c. 
Values for Empty target where scaled so that the integral over $v_z>-450$~cm is the same for Full and Empty setup. The enhancement around $-580$~cm (center of the liquid hydrogen target) corresponds to $p+p$ interactions. (Uncalibrated data, small fraction of statistics, only to illustrate the method) }
\label{vz_distr}
\end{figure}

Now, it is crucial to obtain the value of the $\alpha$ parameter. 
One of the possible methods is to analyze the distribution of the interaction vertex position $v_z$ along the beam axis. 
If one defines $c$ as a ratio of Non-Target events within Full and Empty data samples, 
the $\alpha$ parameter can be expressed using $c$ and numbers of events in Full and Empty data samples.
\begin{eqnarray}
c =  \frac{n^{F}_{NT}}{n^{E}_{NT}}\;\;\;\;\;\; \rightarrow \;\;\;\;\;\; \alpha =  \frac{n^F -  c \cdot n^E}{n^F}
\end{eqnarray}
Estimation of $c$ is possible if we assume that the distribution of $v_z$ in a given region far from the target is due to non target events and is identical for both Full and Empty target events (see Fig. \ref{vz_distr}).
The ratio of the number of events of both data samples in this region is an approximation of $c$
 \begin{eqnarray}
c \approx \left[ \frac{n^{F} }{ n^{E}} \right] _{v_z>-450cm}
\end{eqnarray}
where $v_z>-450$~cm (the TPC gas region) describes the suggested $v_z$ range for the NA61 experiment.

\section{Fluctuation measures}

The scaled variance $\omega$ is used to describe event-by-event fluctuations of the number of produced particles N \cite{omega}
\begin{eqnarray}
\omega = \frac{ \langle N^2 \rangle - {\langle N \rangle}^2 }{ \langle N \rangle }
\label{omega_def}
\end{eqnarray}
To correct $\omega$ for Non-Target interactions one needs to calculate separately 
\begin{eqnarray}
\langle N \rangle_T =  \frac{1}{\alpha} \; [\langle N \rangle^F -  (1-\alpha) \langle N \rangle^E]  \;\;\;\;\; ; 
\;\;\;\;\; \langle N^2 \rangle_T =  \frac{1}{\alpha} \; [\langle {N^2} \rangle^F -  (1-\alpha) \langle {N^2} \rangle^E] ,
\end{eqnarray}
where $\langle N \rangle ^F$,$\langle N^2 \rangle ^F$, $\langle N \rangle ^E$, and $\langle N^2 \rangle ^E$ are quantities calculated independently for Full and Empty data (in the traditional way). The results are then inserted into (\ref{omega_def}):
\begin{eqnarray}
\omega_T = \frac{ \langle N^2 \rangle_T - {\langle N \rangle_T}^2 }{ \langle N \rangle_T }
\end{eqnarray}

The $\Phi$ measure \cite{Phi} is used to investigate event-by-event fluctuations of various per-particle quantities $x$, 
e.g. charge ($\Phi_{q}$), transverse momentum ($\Phi_{p_T}$) or azimuthal angle ($\Phi_{\phi}$).
Although its formal definition contains quantities that are not event mean values as in equations (\ref{emTF}, \ref{emT}), 
the $\Phi$ measure can also be expressed \cite{Liu,Mrow_phi_mod} through
\begin{eqnarray}
\Phi \equiv \sqrt{ 
\frac{\langle X^{2} \rangle}{\langle N \rangle }  - 
\frac{2  \langle X \rangle \; \langle N\; X \rangle}{ {\langle N \rangle}^2 }+
\frac{  {\langle X \rangle}^2 \;   \langle N^{2} \rangle}{ {\langle N \rangle }^3 } }-
\sqrt{ \frac{\langle X_{2} \rangle}{{\langle N \rangle}} - \frac{ {\langle X \rangle}^2 }{  {\langle N \rangle}^2  }}
\label{Phi}
\end{eqnarray}
where $\;\;\;X \equiv \sum_{i=1}^{N}x_i\;\;\;$ and $\;\;\;X_2 \equiv \sum_{i=1}^{N}x_i^2$ are sums over particles in a given event. 
In this equation every quantity within $\langle \; \rangle$ can be calculated for Target interactions using equation (\ref{emT}). Two examples:
\begin{eqnarray}
\langle X \rangle_T =  \frac{1}{\alpha} \; [\langle X \rangle^F -  (1-\alpha) \langle X \rangle^E]  \;\;\;\;\; ; 
\;\;\;\;\; \langle NX \rangle_T =  \frac{1}{\alpha} \; [\langle {NX} \rangle^F -  (1-\alpha) \langle {NX} \rangle^E] ,
\end{eqnarray}
Using such values one can calculate $\Phi$ for Target interactions, by inserting them into equation (\ref{Phi}).

\section{NA61 data}
The above procedure was applied to NA61 $p+p$ data, recorded in 2009. As the data calibration is still in progress we used a small fraction of the available statistics and the results were presented at the conference {\it only} to show the method. The first glimpse of uncalibrated data suggests that the correction due to contamination of Non-Target interactions will be small. For the scaled variance $\omega$ of the multiplicity distributions the absolute value of the difference between $\omega_T$ and $\omega^F$ (fluctuations obtained for Full sample only) divided by $\omega_T$ is lower than 1\%. 
Preliminary NA61 results on the energy dependence of both transverse momentum and multiplicity fluctuations will be available soon.




\clearpage

\clearpage

\begin{center}
FIGURE CAPTIONS
\end{center}

\begin{enumerate}
\item Two-dimensional scan of the phase diagram at SPS energies. Estimated (NA49 results Ref.~\cite{beccat}) and expected (NA61) chemical freeze-out points.
\item Distributions of vertex z position for $p+p$ collisions at beam momentum 40 GeV/c. 
Values for Empty target where scaled so that the integral over $v_z>-450$~cm is the same for Full and Empty setup. The enhancement around $-580$~cm (center of the liquid hydrogen target) corresponds to $p+p$ interactions. (Uncalibrated data, small fraction of statistics, only to illustrate the method)

\end{enumerate}

\end{document}